\documentclass[aps,prx,twocolumn,nofootinbib,groupedaddress,superscriptaddress,longbibliography]{revtex4-1}

\usepackage{amsmath,amssymb}
\usepackage{graphicx}
\usepackage{multirow}
\usepackage{bm}
\usepackage{mathtools}
\usepackage{dsfont}
\usepackage{amsfonts}
\usepackage{xcolor}
\usepackage{ulem}
\usepackage{url}
\usepackage{setspace}

\usepackage[colorlinks=true,linkcolor=magenta,citecolor=magenta]{hyperref}

\def\ba#1\ea{\begin{align}#1\end{align}}
\def\bg#1\eg{\begin{gather}#1\end{gather}}
\def\bpm{\begin{pmatrix}}
\def\epm{\end{pmatrix}}

\renewcommand{\b}[1]{{\boldsymbol #1}}

\newcommand{\bk}{\b k}
\newcommand{\bR}{\b R}

\newcommand{\ket}[1]{| #1 \rangle}

\allowdisplaybreaks

\begin{document}
\title{\textbf{General construction scheme for geometrically nontrivial flat band models}}

\author{Hyeongseop \surname{Kim}}
\altaffiliation{These authors contributed equally to this work}
\affiliation{Department of Physics, Ajou University, Suwon 16499, Korea}

\author{Chang-geun \surname{Oh}}
\altaffiliation{These authors contributed equally to this work}
\affiliation{Department of Applied Physics, University of Tokyo, Tokyo 113-8656, Japan}

\author{Jun-Won \surname{Rhim}}
\email{jwrhim@ajou.ac.kr}
\affiliation{Department of Physics, Ajou University, Suwon 16499, Korea}
\affiliation{Research Center for Novel Epitaxial Quantum Architectures, Department of Physics, Seoul National University, Seoul, 08826, Korea}

\begin{abstract}
A singular flat band(SFB), a distinct class of the flat band, has been shown to exhibit various intriguing material properties characterized by a geometric quantity of the Bloch wave function called the quantum distance.
We present a general construction scheme for a tight-binding model hosting an SFB, where the quantum distance profile can be controlled.
We first introduce how to build a compact localized state(CLS), a characteristic eigenstate of the flat band, providing the flat band with a band-touching point, where a specific value of the maximum quantum distance is assigned.
Then, we develop a scheme designing a tight-binding Hamiltonian hosting an SFB starting from the obtained CLS, satisfying the desired hopping range and symmetries by applying the construction scheme.
While the scheme can be applied to any dimensions and lattice structures, we propose several simple SFB models on the square and kagome lattices. 
%
Finally, we establish a bulk-boundary correspondence between the maximum quantum distance and the boundary modes for the open boundary condition, which can be used to detect the quantum distance via the electronic structure of the boundary states. 
\end{abstract}

\maketitle

\section{Introduction \label{sec: Intro}}
When a band has a macroscopic degeneracy, we call it a flat band~\cite{leykam2018artificial,rhim2021singular}.
Flat band systems have received great attention because their van Hove singularity is expected to stabilize various many-body states when the Coulomb interaction is introduced.
Examples of such correlated states induced by flat bands are unconventional superconductivity~\cite{volovik1994fermi,cao2018unconventional,liu2021spectroscopy,balents2020superconductivity,peri2021fragile,yudin2014fermi,volovik2018graphite,aoki2020theoretical,kononov2021superconductivity}, ferromagnetism~\cite{mielke1993ferromagnetism,tasaki1998nagaoka,mielke1999stability,hase2018possibility,you2019flat,saito2021hofstadter,sharpe2019emergent}, Wigner crystal~\cite{wu2007flat,chen2018ferromagnetism,jaworowski2018wigner}, and fractional Chern insulator~\cite{wang2011nearly,tang2011high,sun2011nearly,neupert2011fractional,sheng2011fractional,regnault2011fractional,weeks2012flat,yang2012topological,liu2012fractional,bergholtz2013topological}.
%
%
Recently, it was revealed that the flat band could be nontrivial from the perspective of geometric notions, such as the quantum distance, quantum metric, and cross-gap Berry connection~\cite{rhim2020quantum,hwang2021geometric,peotta2015superfluidity,torma2022superconductivity,piechon2016geometric}.
The quantum distance is related to the resemblance between two quantum states defined by
\ba
d^2 = 1 - |\langle \psi_1 | \psi_2 \rangle |^2,\label{eq:quantum_distance}
\ea
which is positive-valued and ranging from 0 to 1~\cite{buvzek1996quantum,dodonov2000hilbert,wilczek1989geometric}.
If a flat band has a band-touching point with another parabolic band and the maximum value of the quantum distance, denoted by $d_\mathrm{max}$, between eigenvectors around the touching point is nonzero, we call it a singular flat band(SFB)~\cite{rhim2019classification}.
The singular flat band hosts non-contractible loop states featuring exotic topological properties in real space~\cite{bergman2008band,ma2020direct}.
The Landau level structure of the singular flat band is shown to be anomalously spread into the band gap region~\cite{rhim2020quantum,hwang2021geometric}, and the maximum quantum distance determines the magnitude of the Landau level spreading.
Moreover, if we introduce an interface in the middle of a singular flat band system by applying different electric potentials, an interface mode always appears, and the maximum quantum distance determines its effective mass~\cite{oh2022bulk}

Diverse unconventional phenomena characterized by quantum distance are expected to occur in the singular flat band systems. 
However, we lack good tight-binding models hosting the singular flat band where one can control the quantum distance, although numerous flat band construction methods have been developed~\cite{PhysRevB.99.235118,cualuguaru2022general,mizoguchi2020systematic,graf2021designing,PhysRevB.103.L241102,hwang2021flat,hwang2021general,maimaiti2019universal,huda2020designer}.
This paper suggests a general construction scheme for the tight-binding Hamiltonians with a singular flat band and the controllable maximum quantum distance.
The construction process's essential part is designing a compact localized state(CLS), which gives the desired maximum quantum distance.
The CLS is a characteristic eigenstate of the flat band, which has finite amplitudes only inside a finite region in real space~\cite{rhim2019classification}.
The CLS can be transformed into the Bloch eigenstate, and any Hamiltonian having this as one of the eigenstates must host a flat band~\cite{rhim2019classification}.
Among infinitely many possible tight-binding Hamiltonians for a given CLS, one can choose several ones by implementing the wanted symmetries and hopping range into the construction scheme.
Using the construction scheme, we suggest several simple tight-binding models hosting a singular flat band and characterized by the maximum quantum distance on the square and kagome lattices.
Using the obtained tight-binding models, we propose a bulk-boundary correspondence of the flat band system from the maximum quantum distance to address a question of how to measure the maximum quantum distance in experiments.
The previous work established the bulk-interface correspondence for the interface between two domains with different electric potentials in the same singular flat band system, where the maximum quantum distance of the bulk determines the interface mode's effective mass~\cite{oh2022bulk}.
We show that the same correspondence applies to open boundaries if a boundary mode exists.

The paper is organized as follows.
In Sec.~\ref{sec: Scheme}, we introduce a general flat band construction scheme, which starts from a given CLS.
In Sec.~\ref{sec:dmax}, we present how to construct a CLS characterized by a desired value of $d_\mathrm{max}$.
Combining these two methods, we build two tight-binding models hosting a singular flat band characterized by $d_\mathrm{max}$ in the kagome lattice and square lattice bilayer in Sec.~\ref{sec:TB_models}.
Then, in Sec.~\ref{sec:bbc}, we propose the bulk-boundary correspondence characterized by the quantum distance.
Finally, we summarize and discuss our results in Sec.~\ref{sec:conclusions}.

\section{General flat band construction scheme} \label{sec: Scheme}

Since the key ingredient of the flat band construction scheme is designing a CLS, we begin with a brief review of it.
The general form of the Bloch wave function of the $n$-th band with momentum $\bk$ is given by
\bg
\ket{\psi_{n,\bk}}=\frac{1}{\sqrt{N}}\sum_{\bR}\sum_{q=1}^{Q}e^{i\bk\cdot\bR} v_{n,\bk,q}\,\ket{\bR,q},
\label{eq:Bloch_eigenfunction}
\eg
where $N$ is the number of unit cells in the system, $\bR$ represents the position vectors of the unit cells, $\ket{\bR,q}$ corresponds to the $q$-th orbital among $Q$ orbitals in a unit cell, and $v_{n,\bk,q}$ is the $q$-th component of the eigenvector $\mathbf{v}_{n,\bk}$ of the $Q\times Q$ Bloch Hamiltonian~\cite{rhim2017bulk}.
Then it was shown that if the $n_0$-th band is flat, one can always find a linear combination of the Bloch wave functions resulting in the CLS of the form:
\ba
\ket{\chi_{\bR}}=c_{\chi}\sum_{\bk\in{\rm BZ}}\sum_{\bR^\prime} \sum_{q=1}^{Q} \alpha_{\bk} v_{n_0,\bk,q} e^{i\bk\cdot(\bR^\prime -\bR)} \ket{\bR^\prime,q},
\label{eq: CLS}
\ea
where $c_{\chi}$ is the normalization constant and $\alpha_{\bk}$ is a mixing coefficient of the linear combination~\cite{rhim2019classification}.
It is important to note that $\alpha_{\bk}v_{n_0,\bk,q}$ is a finite sum of exponential factors $e^{i\bk\cdot\bR}$ so that the range of $\bR^\prime$ in (\ref{eq: CLS}) with the nonzero coefficient of $\ket{\bR^\prime,q}$ is finite.
If $\alpha_{\bk}v_{n_0,\bk,q} = 0$ at $\bk=\bk_0$ for all kinds of $\alpha_{\bk}$ satisfying the above properties, we call the band the singular flat band because $v_{n_0,\bk,q}$ becomes discontinuous at $\bk_0$ in this case.
From (\ref{eq: CLS}), one can note that the constants in front of each exponential factor of $\alpha_{\bk}v_{n_0,\bk,q}$ becomes the amplitude of the CLS.

\begin{figure}[htb]
\centering
\includegraphics[width=0.48\textwidth ]{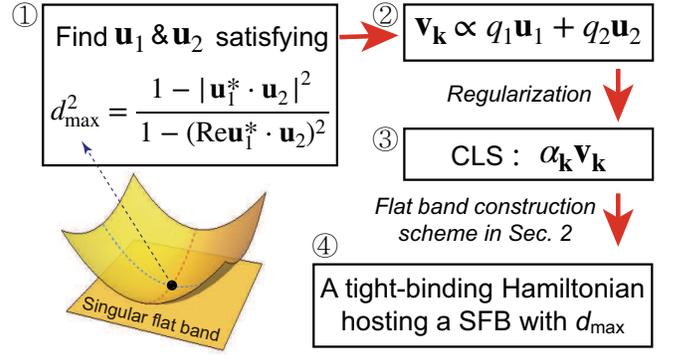}
\caption{A scheme for the construction of a tight-binding model hosting a singular flat band(SFB) characterized by the maximum quantum distance($d_\mathrm{max}$). First, we find two vectors with complex components to yield the desired $d_\mathrm{max}$. Then, we build a CLS from the two vectors in the second and third steps. Finally, we obtain an SFB Hamiltonian from the CLS using the general flat band construction scheme given in Sec.~\ref{sec: Scheme}.
}\label{fig:scheme}
\end{figure}

We construct a flat band Hamiltonian from a CLS arbitrarily designed on a given lattice.
This part corresponds to the third and fourth stages of the construction scheme sketched in Fig.~\ref{fig:scheme}.
By using the correspondence between the CLS and Bloch eigenvector in (\ref{eq: CLS}), one can obtain $\alpha_{\bk}v_{n_0,\bk,q}$ in the form of the finite sum of exponential factors from the designed CLS.
Then, by normalizing $\alpha_{\bk}v_{n_0,\bk,q}$, one can have the flat band's eigenvector $v_{n_0,\bk,q}$ corresponding to the CLS.
Our purpose is to find a tight-binding Hamiltonian of the form
\ba
H_{ij}^{\rm lattice}(\b{k})=\sum_{\Delta\b{R}}t_{ij}(\Delta\b{R})e^{-i\b{k}\cdot\Delta\b{R}},
\label{eq: Hamiltonian matrix}
\ea
which satisfies
\ba
\left[ H_{ij}^{\rm lattice}(\b{k}) - E_\mathrm{flat} \right] \alpha_{\bk}\mathbf{v}_{n_0,\bk} =0,\label{eq:flat_band_equation}
\ea
where $E_\mathrm{flat}$ is the flat band's energy and $\mathbf{v}_{n_0,\bk}$ is a column vector with components $v_{n_0,\bk,q}$.
Here, $t_{ij}(\Delta\b{R})$ represents the hopping parameter between the $i$-the and $j$-th orbitals in unit cells separated by $\Delta\bR = \sum_{\nu=1}^d n_\nu \mathbf{a}_\nu$, where $n_\nu$ is an integer, $d$ is spatial dimension, and $\mathbf{a}_\nu$ is the primitive vector.
For convenience, we denote $t_{ij}^{n_{1}, n_{2}\dots n_{\nu}} \equiv t_{ij}(\Delta\bR)$ and $e_{\nu} \equiv e^{-i\bk\cdot\b a_{\nu}}$.
We use a bar notation for the complex conjugate such that $\overline{t_{ij}^{n_{1}, n_{2}\dots n_{\nu}}} = (t_{ij}^{n_{1}, n_{2}\dots n_{\nu}})^*$ and $\overline{e_{\nu}} = (e_{\nu})^*$.
Then, the matrix element of the tight-binding Hamiltonian is rewritten as
\ba
H_{ij}^{\rm Lattice}(\b{k})=\sum_{n_{1},n_{2}\dots n_{\nu}}\sum_{ij}t_{i j}^{n_{1}, n_{2}\dots n_{\nu}}\prod_{\nu^\prime}e_{\nu^\prime}^{n_{\nu^\prime}}.
\label{eq: general Hamiltonian matrix}
\ea
Here, the hopping parameters $t_{i j}^{n_{1}, n_{2}\dots n_{\nu}}$ can be considered complex unknowns determined by the matrix equation in (\ref{eq:flat_band_equation}).
One can encode some wanted hopping range and symmetries by manipulating the number of unknown hopping parameters and setting relations between them, respectively.
Noting that $\alpha_{\bk}\mathbf{v}_{n_0,\bk} = \sum_{n_1,n_2,\cdots,n_\nu} c_{n_1,n_2,\cdots,n_\nu} \prod_{\nu^\prime}e_{\nu^\prime}^{n_{\nu^\prime}}$ as described above, the matrix equation (\ref{eq:flat_band_equation}) leads to a system of linear equations obtained from the coefficients of the independent exponential factors.
%

\begin{figure}[htb]
\centering
\includegraphics[width=0.45\textwidth ]{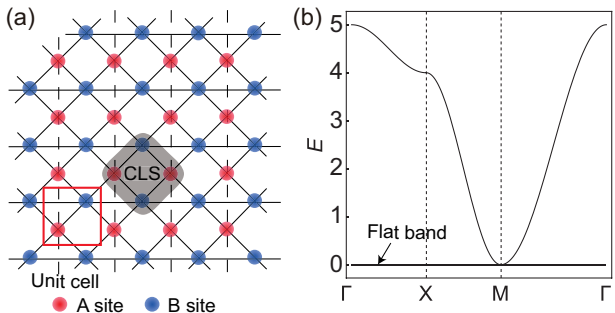}
\caption{(a) The checkerboard flat band model with $d_\mathrm{max}=1$, denoted by CB$_1$. A red box represents the unit cell. The hopping amplitudes are 1 for the dashed lines along the $y$-axis, $-a/b$ for black solid lines along diagonal directions, and $a^2/b^2$ for the blue solid lines along the $x$-axis. The CLS corresponding to the flat band is drawn by a gray region. The CLS's amplitudes are $a$ at the A-sites and $b$ at the B-sites. (b) The band structure of the checkerboard model for $a=1$ and $b=2$.
}\label{fig:cb1}
\end{figure}
Let us consider a simple example, the flat band Hamiltonian on the checkerboard lattice, which is illustrated in Fig.~\ref{fig:cb1}(a).
We design a CLS in the shape of a square represented by a gray region in Fig.~\ref{fig:cb1}(a), having amplitudes $a$ and $b$ on the A and B sites, respectively.
From the CLS, one can obtain the flat band's eigenvector $\alpha_{\bk}\mathbf{v}_{n_0,\bk}$ in momentum space such that the CLS's amplitude in the unit cell $\Delta\bR = \sum_{\nu=1}^d n_\nu \mathbf{a}_\nu$ becomes the coefficient of the exponential factor $\prod_\nu e^{\nu}_\nu$.
As a result, we have
\ba
\alpha_{\bk}\mathbf{v}_{n_0,\bk} = \begin{pmatrix}
a+a e_1 \\ b+ b \overline{e_2}
\end{pmatrix}.\label{eq:cb_cls}
\ea
The next step is to design the tight-binding Hamiltonian (\ref{eq: general Hamiltonian matrix}).
We seek one with real-valued hopping parameters up to the next-nearest hopping range.
Then, the matrix elements of $H^{{\rm CB}_1}$ are of the form 
\begin{align}
    H_{11}^{{\rm CB}_1} &= t_{11}^{0,0} + t_{11}^{0,-1} \overline{e_2} +t_{11}^{0,1} e_2,\\
    H_{12}^{{\rm CB}_1} &= t_{12}^{0,0} + t_{12}^{1,0} e_1 + t_{12}^{0,1} e_2 +t_{12}^{1,1} e_1 e_2,\\
    H_{22}^{{\rm CB}_1} &= t_{22}^{0,0} + t_{22}^{-1,0} \overline{e_1} +t_{22}^{1,0} e_1,
\end{align}
%
From the flat band condition (\ref{eq:flat_band_equation}) and by enforcing the hermicity, one can find relationships between the tight-binding parameters, which lead to the following form of the Hamiltonian:
\ba
H^{{\rm CB}_1} = \begin{pmatrix}
    -2(1+\cos k_y) & \frac{a}{b}(1+e_1)(1+e_2) \\ \frac{a}{b}(1+\overline{e_1})(1+\overline{e_2}) & -\frac{2a^2}{b^2}(1+\cos k_x)
\end{pmatrix},
\ea
where we further assume that $a$ and $b$ are real constants and $t_{11}^{0,0}=-2$ for convenience.
This Hamiltonian yields a zero-energy flat band and lower parabolic with a singular band-touching point at $\mathbf{k} = (\pi,\pi)$ as plotted in Fig.~\ref{fig:cb1}(b).
In fact, this band-crossing is already designed at the construction stage of the CLS in (\ref{eq:cb_cls}) by assigning a simultaneous zero of all the components of $\alpha_{\bk}\mathbf{v}_{n_0,\bk}$ at $\mathbf{k} = (\pi,\pi)$.

\begin{figure}[htb]
\centering
\includegraphics[width=0.45\textwidth ]{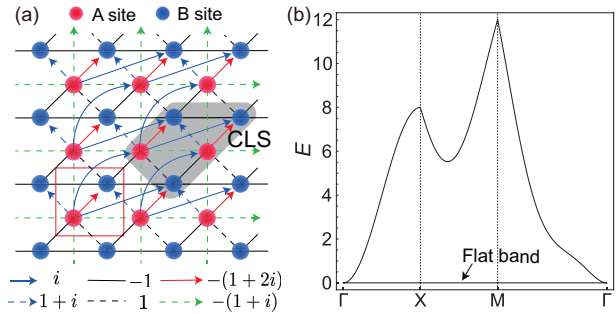}
\caption{(a) The checkerboard flat band model with $d_\mathrm{max}=1/\sqrt{2}$, denoted by CB$_2$. A red box represents the unit cell. The hopping parameters are given below the figure. For the complex hopping processes, the hopping direction is represented by the arrow. (b) The band structure of the checkerboard model CB$_2$.
}\label{fig:cb2}
\end{figure}

\section{Maximum quantum distance}\label{sec:dmax}
In this section, we discuss how to endow the band-crossing of the flat band with the wanted value of the maximum quantum distance $d_\mathrm{max}$ when we construct a flat band model.
Specifically, the quantum distance between two Bloch eigenstates with momenta $\bk$ and $\bk^\prime$ is denoted as $d(\bk,\bk^{\prime})^2= 1 - |\mathbf{v}_{\bk^\prime}^*\cdot \mathbf{v}_\bk|^2$ and $d_\mathrm{max}$ is defined as
\ba
    d_{\rm max}^{2}=\lim_{r_{D}\rightarrow 0}\max{[d(\bk,\bk^{\prime})^{2}]}\Bigm|_{\bk,\bk^{\prime}\in D(\bk_{0})},\label{eq: maximum quantum distance}
\ea
where $\mathbf{v}_\bk$ is the flat band's eigenvector and $D(\bk_{0})$ is a closed disk with radius $r_{D}$ centered at the band-crossing point $\bk_{0}$~\cite{rhim2020quantum}.
In the previous study, $d_\mathrm{max}$ was proposed to measure the strength of the singularity at $\bk_{0}$.
Note that if there is a singularity at $\bk_{0}$, the quantum distance between the Bloch eigenstates can remain finite even if the momenta of them are very close to $\bk_{0}$~\cite{rhim2020quantum}.
For the well-known singular flat band models, such as the kagome and checkerboard lattice models, $d_\mathrm{max}$ is found to be unity.
While $0 \leq d_\mathrm{max} \leq 1$ in general~\cite{rhim2020quantum}, there have been almost no examples of the tight-binding models hosting $d_\mathrm{max}$ smaller than 1.

One can design $\mathbf{v}_\bk$ of the flat band to have a specific value of $d_\mathrm{max}$ by manipulating the form of the linear expansion of $\alpha_{\bk}\mathbf{v}_{\bk}$ around the band-crossing point.
Denoting $q_{\mu} = k_{\mu}-k_{0,\mu}$, where $\bk_0$ is the band-crossing point, the eigenvector $\alpha_{\bk}\mathbf{v}_{\bk}$ can be written as
\ba
\alpha_{\bk}\mathbf{v}_{\bk} \simeq q_{1}\mathbf{u}_1 + q_{2}\mathbf{u}_2,\label{eq:linear_expansion}
\ea
in the vicinity of $\bk_0$ up to the linear order of $\mathbf{q}$.
Here, $\mathbf{u}_1$ and $\mathbf{u}_2$ are $Q\times 1$ constant normalized vectors.
Then, one can show that
\ba
d_{\rm max}^{2}=\frac{1-|\mathbf{u}_1^*\cdot\mathbf{u}_2|^{2}}{1-({\rm Re}\mathbf{u}_1^*\cdot\mathbf{u}_2)^{2}}.\label{eq: dmax}
\ea
See Appendix~\ref{app: dmax} for the detailed derivations.
By using this relationship, one can choose two constant vectors $\mathbf{u}_1$ and $\mathbf{u}_2$, giving the desired value of $d_\mathrm{max}$.
Then, performing a regularization of (\ref{eq:linear_expansion}) by applying transformations, such as $q_i \rightarrow \sin q_i$ and $q_i \rightarrow 1-e^{iq_i}$, one can obtain $\alpha_{\bk}\mathbf{v}_{\bk}$, the Fourier transform of a CLS, in the form of a finite sum of exponential factors $e_\nu$ and $\overline{e_{\nu}}$.
In this stage, corresponding to the first to third steps in Fig.~\ref{fig:scheme}, one can control the size of the CLS, which is closely related to the hopping range of the tight-binding model obtained from this CLS.
Once we obtain $\alpha_{\bk}\mathbf{v}_{\bk}$, the tight-binding Hamiltonian with the desired $d_\mathrm{max}$ can be built by using the construction scheme in the previous section.

From the $d_\mathrm{max}$-formula (\ref{eq: dmax}), one can note that $d_\mathrm{max}$ can be less than one and larger than zero only when $\mathbf{u}_1^*\cdot\mathbf{u}_2$ is not real or pure imaginary.
Namely, $u_{1,m}^*u_{2,m}$ should be imaginary at least for one $m$, where $u_{i,m}$ is the $m$-th component of $\mathbf{u}_i$.
Let us denote such an index $m$ by $m_0$.
Then, the $m_0$-th component of $\alpha_\mathbf{k}\mathbf{v}_\mathbf{k}$, given by $\alpha_\mathbf{k}\mathbf{v}_\mathbf{k}|_{m_0} = u_{1,m_0}q_1 + u_{2,m_0}q_2$, must be regularized into a form, where the coefficients of the exponential factors contain both the real and imaginary values.
This implies that the CLS corresponding to the singular flat band with $0<d_\mathrm{max} <1$ cannot be constructed only with the real amplitudes.
Note that the CLS of the flat band of the kagome lattice can be represented by only real amplitudes because the corresponding $d_\mathrm{max}$ is unity.
However, the CLS should consist of different complex amplitudes in at least two atomic sites for generic flat bands with $0<d_\mathrm{max} <1$.
The tight-binding Hamiltonian stabilizing such a CLS usually requires complex hopping parameters.
Moreover, it is shown in Appendix \ref{app:dmax_cond} that we need more that two exponential factors for at least one component of $\alpha_{\bk}\mathbf{v}_{\bk}$.
This implies that we usually need hopping processes between atoms at a longer distance than the nearest neighbor ones.

Let us consider the checkerboard lattice example again.
We assume that the touching point is at $\bk=(0,0)$. 
First, to obtain a model with $d_\mathrm{max}=1$, we can choose $\mathbf{u}_1= (i, 0 )^\mathrm{T}$ and $\mathbf{u}_2= ( 0 , -i )^\mathrm{T}$ in (\ref{eq:linear_expansion}), using the formula (\ref{eq: dmax}).
Then, we apply the regularization $ik_1 \rightarrow 1-e^{-i k_1}$ and $ik_2 \rightarrow 1-e^{i k_2}$ to obtain the CLS's Fourier transform.
Second, on the other hand, one can let the CLS have $d_\mathrm{max}=1/\sqrt{2}$ by choosing $\mathbf{u}_1= ( i , -1 )^\mathrm{T}/\sqrt{2}$ and $\mathbf{u}_2= ( 0 , -i )^\mathrm{T}$.
In this case, an example of the regularization gives $\alpha_{\bk}\mathbf{v}_{\bk} = ( 1-e^{-i k_1} , 1+i-i e^{-i k_{1}}-e^{i k_{2}} )^\mathrm{T}$.
The CLS corresponding to this eigenvector is drawn in Fig.~\ref{fig:cb2}(a).
An example of the flat band tight-binding Hamiltonian obtained from this choice of the CLS is given by
\begin{align}
    H^{\mathrm{CB}_2} = \begin{pmatrix}
        v_{2}v_{2}^{*} && -v_{1}v_{2}^{*} \\
-v_{2}v_{1}^{*} && v_{1}v_{1}^{*}
    \end{pmatrix},
\end{align}
where $v_{1}=1-e^{-i k_{1}}$ and $v_{2}=1+i-ie^{-i k_{1}}-e^{i k_{2}}$.
The band structure of this model is shown in Fig.~\ref{fig:cb2}(b).
One can note that the band has non-zero slopes at X and M points due to the broken time-reversal, mirror, and inversion symmetries.
As discussed above, the CLS contains both the real and imaginary amplitudes and the Hamiltonian possesses imaginary hopping processes in the $d_\mathrm{max}=1/\sqrt{2}$ case.

\begin{figure}[htb]
\centering
\includegraphics[width=0.45\textwidth ]{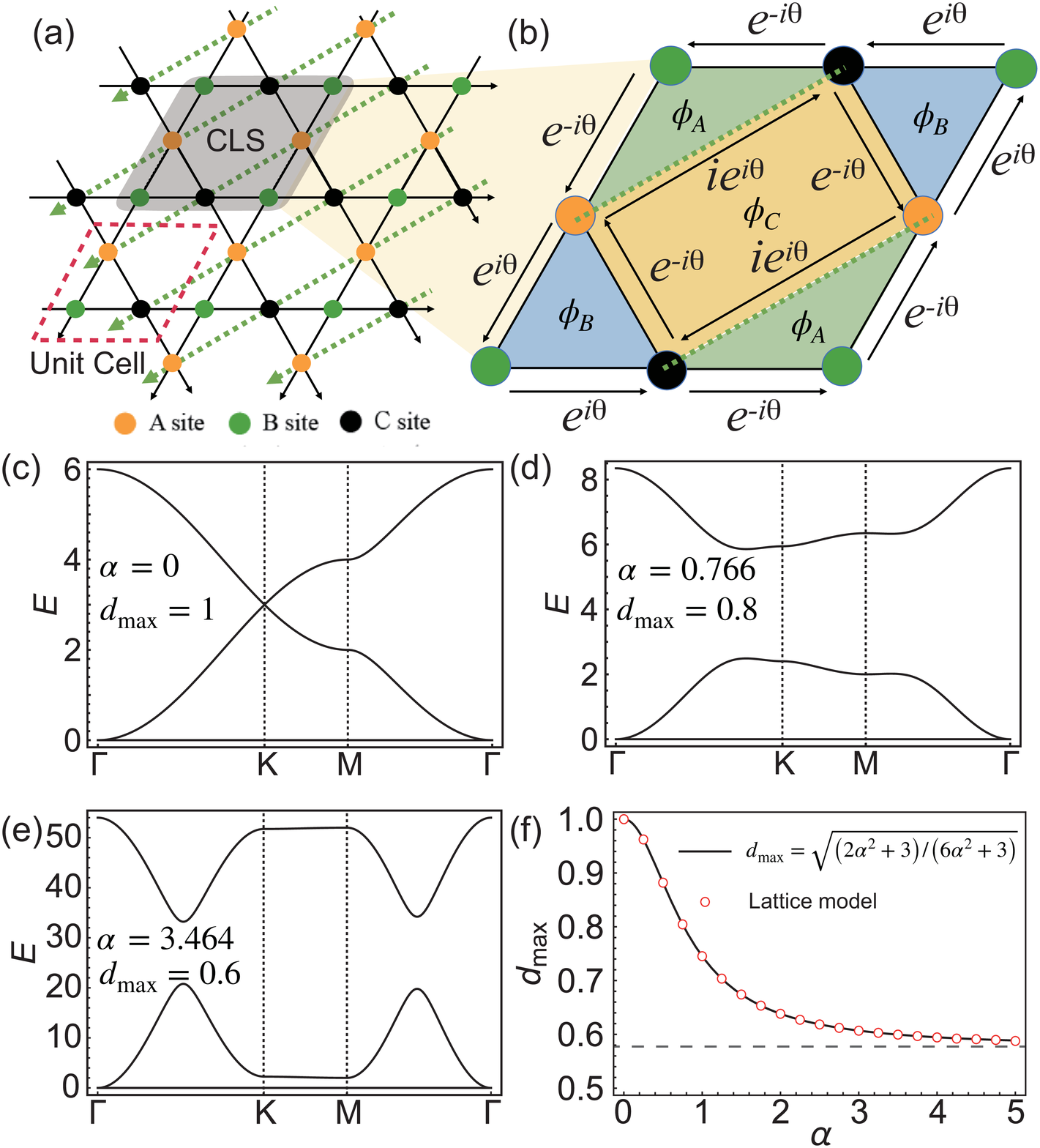}
\caption{The kagome lattice model hosting a flat band characterized by the quantum distance. (a) The nearest and the next nearest neighbor hopping processes are denoted by the black solid and green dashed lines, respectively. The CLS corresponding to the flat band of this model is represented by the gray region. (b) The phase parts of the hopping parameters are highlighted. The magnetic fluxes for the complex hopping parameters are given by $\phi_A=\pi/2-\theta$, $\phi_B=\theta$, and $\phi_C=-\pi$. (c-e) Band dispersions for $\alpha=0$, $\alpha=0.766$, and $\alpha=3.464$. (f) $d_\mathrm{max}$ as a function of $\alpha$. The formula (\ref{eq:kagome_dmax}) drawn by a black curve is compared with the numerically calculated $d_\mathrm{max}$ from the lattice model, represented by circles.
}\label{fig:kagome}
\end{figure}

\section{Flat band models characterized by the quantum distance} \label{sec:TB_models}

\subsection{Kagome lattice model}
We construct a simple tight-binding model hosting a SFB characterized by $d_\mathrm{max}$ in the kagome lattice.
When we consider only the nearest neighbor hopping processes in the kagome lattice, which is the most popular case, the flat band already has a quadratic band-touching, but the corresponding $d_\mathrm{max}$ is fixed to 1~\cite{rhim2020quantum}.
We generalize this conventional kagome lattice model so that $d_\mathrm{max}$ can vary by adding some next-nearest neighbor hopping processes.

\begin{figure}[htb]
\centering
\includegraphics[width=0.45\textwidth ]{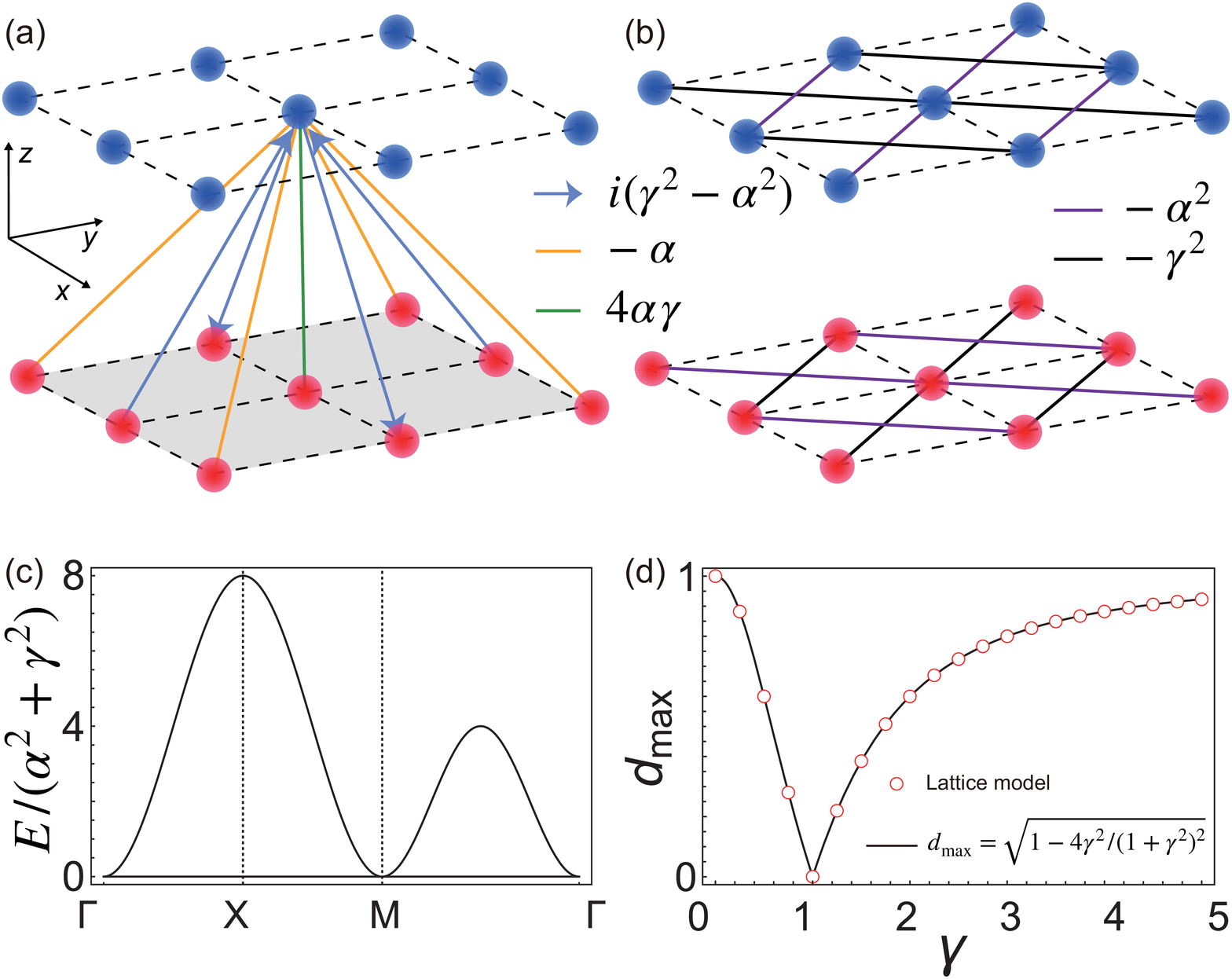}
\caption{The square lattice bilayer model hosting a flat band characterized by the quantum distance. We plot the interlayer and intralayer hopping processes in (a) and (b), respectively. In (c), we plot the band structure, where the energy is scaled by $\alpha^2+\gamma^2$. The relation between $d_\mathrm{max}$ and the band parameter $\gamma$ is presented in (d).
}\label{fig:bilayer}
\end{figure}
We begin with two vectors $\mathbf{u}_1=c_1(-i,-2\alpha,-i)^\mathrm{T}$ and $\mathbf{u}_2=c_2(0,-i-\alpha,-i)^\mathrm{T}$, where $c_1=(2+4\alpha^2)^{1/2}$ and $c_2=(2+\alpha^2)^{1/2}$.
This set of vectors yields
\begin{align}
    d_\mathrm{max} = \sqrt{\frac{3+2\alpha^2}{3+6\alpha^2}},\label{eq:kagome_dmax}
\end{align}
where $\alpha$ can take any real number from $-\infty$ to $\infty$.
As shown in Fig.~\ref{fig:kagome}(f), $d_\mathrm{max}$ of the constructed SFB model can take values from $1/\sqrt{3}$ to 1.
Then, we regularize the linearized vector $\mathbf{v}_\mathrm{fb}=\mathbf{u}_1 k_1 + \mathbf{u}_2 k_2$ to
\begin{align}
\mathbf{v}_\mathrm{fb}= \bpm
1-\overline{e_{1}} \\
-1+i\alpha \overline{e_{1}}+e_{2}-i\alpha e_{3} \\
e_{1}-\overline{e_{2}}
\epm,
\end{align}
where $e_{3}=e_1 e_2$.
The CLS corresponding to this eigenvector of the flat band is drawn in Fig.~\ref{fig:kagome}(a) by the gray region.
From this choice of the CLS, we construct a tight-binding Hamiltonian as follows:
\ba
H_{\rm kag}(\bk)=
\bpm
g_{1} && g_{2}^{*} && g_{3}^{*} \\
g_{2} && 2 && g_{4}^{*} \\
g_{3} && g_{4} && g_{1}
\epm,
\ea
where $g_{1} = 2|t|^{2}$, $g_{2} = t(1+\overline{e_{3}})$, $g_{3} = t(1+e_{2} + i\alpha t (\overline{e_{1}}+e_{3})$, $g_{4} = t(1+\overline{e_{1}})$, $t = e^{i\theta}\sqrt{1+\alpha^{2}} $, and $\theta=\cos^{-1}(1/\sqrt{1+\alpha^2})$.
Note that when $\alpha=0$, where $d_\mathrm{max}=1$, the model reduces to the kagome lattice model with only nearest neighbor hopping processes.
As the parameter $\alpha$ grows, the nearest neighbor hopping parameters become complex-valued, and the next nearest neighbor hopping processes are developed as represented by green dashed lines in Fig.~\ref{fig:kagome}(a).
One can assign threading magnetic fluxes corresponding to the complex hopping parameters as illustrated in Fig.~\ref{fig:kagome}(b), similar to the Haldane model in graphene.
In Fig.~\ref{fig:kagome}(c) to (e), we plot band dispersions for various values of $\alpha$, where we have a zero-energy flat band at the bottom.
Fig.~\ref{fig:kagome}(c) is the well-known band diagram of the kagome lattice with the nearest neighbor hopping processes.
If $\alpha$ is nonzero, the Dirac point is gapped out due to the broken $C_6$ symmetry, but the quadratic band-crossing at the $\Gamma$ point is maintained.
We calculate $d_\mathrm{max}$ of this model directly using (\ref{eq: maximum quantum distance}) and check that the continuum formula (\ref{eq:kagome_dmax}) works well as shown in Fig.~\ref{fig:kagome}(f).

\begin{figure*}
\centering
\includegraphics[width=1\textwidth ]{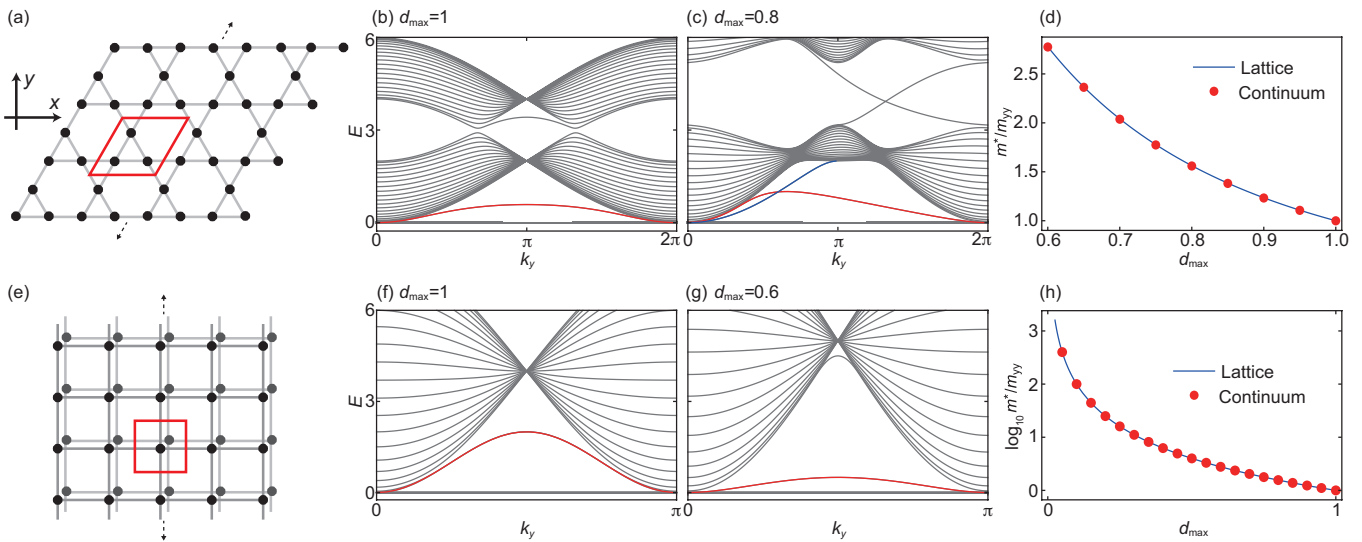}
\caption{The Bulk-boundary correspondence of singular flat band systems. The upper(lower) panels correspond to the results of the kagome lattice(square lattice bilayer) model. (a) and (e) illustrate the lattice structures of the ribbon geometries of the kagome lattice and square lattice bilayer, respectively. The red boxes are the unit cells. The system is terminated along the $x$-axis, and there is a finite number of unit cells, denoted by $W$, along this direction. (b,c) and (f,g) are the band structures of the kagome lattice and square lattice bilayer with $W=20$, respectively. Red and blue lines represent the boundary modes. In (d) and (h), we plot the effective mass of the boundary modes around $k_y=0$ as a function of $d_\mathrm{max}$ and compare it with the continuum result in (\ref{eq:bbc1}).
}\label{fig:bbc}
\end{figure*}

\subsection{Square lattice bilayer model}
We also construct an SFB tight-binding model in the square lattice bilayer, where one can adjust $d_\mathrm{max}$.
The lattice structure is illustrated in Fig.~\ref{fig:bilayer}(a) and (b).
As in the kagome lattice case, the construction scheme starts from setting two constant vectors.
Our choice is $\mathbf{u}_{1}=c(i\alpha+\gamma,-\alpha-i\gamma)^\mathrm{T}$, and $\mathbf{u}_{2}=\overline{\mathbf{u}}_{1}$, where $c=(2\alpha^{2}+2\gamma^{2})^{1/2}$. 
One can show that $d_\mathrm{max}$ calculated from these vectors is given by
\begin{align}
    d_\mathrm{max} = \sqrt{1-\frac{4\alpha^{2}\gamma^{2}}{(\alpha^{2}+\gamma^{2})^{2}}},
\end{align}
where $\alpha$ and $\gamma$ can take any real values.
$d_\mathrm{max}$ can vary from 0 to 1.
If $\alpha$ or $\gamma$ is zero

As shown in Fig.~\ref{fig:bilayer}(d), $d_{\mathrm{max}}$ of the constructed SFB model can take values from 0 to 1. Then, we regularize a vector $\mathbf{v}_{\mathrm{fb}}=\mathbf{u}_{1}k_{1}+\mathbf{u}_{2}k_{2}$ to
\begin{align}
    \mathbf{v}_{\mathrm{fb}} =
    \begin{pmatrix}
        -i\gamma(1 - e_{1}e_{2}) - \alpha (e_{1}-e_{2}) \\
        i\alpha(1 - e_{1}e_{2}) + \gamma(e_{1} - e_{2})
    \end{pmatrix}
    .
\end{align}
The CLS corresponding to this eigenvector of the flat band is drawn in Fig.~\ref{fig:bilayer}(e) by a gray region. From this choice of the CLS, we construct a tight-binding Hamiltonian as fallows:
\ba
H_{\mathrm{sq}}(\bk) =
    \bpm
    |f_{2}|^{2} && f_{3} \\
    \overline{f_{3}} && |f_{1}|^{2}
    \epm
    ,
\ea
where $f_{1} = -i\gamma(1 - e_{1}e_{2}) - \alpha (e_{1}-e_{2})$, $f_{2} = i\alpha(1 - e_{1}e_{2}) + \gamma(e_{1} - e_{2})$ and $f_{3} = -f_{1}\overline{f_{2}}$.
When $\alpha=\gamma$ or $\alpha$ and $\gamma$ are zero, $d_\mathrm{max}=1$.
As parameters $\alpha$ and $\gamma$ grows, interlayer and intralayer hopping appears and if $\alpha \neq \gamma$, complex-value hopping process is developed as represented by blue arrow in Fig.~\ref{fig:bilayer}(a).
Unlike kagome lattice model, this model has an isotropic band dispersion. As show in Fig.~\ref{fig:bilayer}(c), we plot the zero-energy flat band at the bottom and a band dispersion that is independent of variables $\alpha$ and $\gamma$.
Fig.~\ref{fig:bilayer}(d) shows $d_\mathrm{max}$ of this model which is calculated by continuum formula (\ref{eq: dmax}) and directly using (\ref{eq: maximum quantum distance}).

\section{Bulk-boundary correspondence}\label{sec:bbc}
The bulk-boundary correspondence is the essential idea of the topological analysis of materials~\cite{kane2005quantum,kane2005z,bernevig2006quantum,hatsugai1993chern,kitaev2009periodic,fukui2012bulk,mong2011edge,rhim2018unified}.
Based on this, one can detect the topological information of the bulk by probing the electronic structure of the boundary states.
%
%
Recently, a new kind of bulk-interface correspondence from the quantum distance for the flat band systems was developed~\cite{oh2022bulk}.
Here, a specific type of interface is considered, which is generated between two domains of a singular flat band system with different onsite potentials $U_R$ and $U_L$.
Note that the two domains are characterized by the same geometric quantity $d_\mathrm{max}$, unlike the topological bulk-boundary correspondence, where the boundary is formed between two regions with different topologies.
In the case of the singular flat band systems, an interface state is guaranteed to exist if the value of $d_\mathrm{max}$ is nonzero, and the corresponding band dispersion around the band-crossing point is given by
\begin{align}
    E_\mathrm{IF}(k) \approx \frac{d^2_\mathrm{max}}{2m_b} k^2 + U_0,\label{eq:bbc1}
\end{align}
where $k$ and $m_b$ are the crystal momentum and the bulk mass along the direction of the interface, respectively, and $U_0 = \mathrm{min}(U_R,U_L)$.
This formula implies that the effective mass of the interface mode is $m^* = m_b/d_\mathrm{max}^2$.

Now, we examine the formula (\ref{eq:bbc1}) for the finite systems satisfying the open boundary condition.
In the previous work, (\ref{eq:bbc1}) could be obtained by presuming an exponentially decaying edge mode, and the existence of such a state was guaranteed for the specific interface of the step-like potential.
While the open boundaries are naturally induced when we prepare a sample, the application of the step-like potential is not usually straightforward in experiments.
Therefore, it is worthwhile to investigate the bulk-boundary correspondence for the open boundary systems.
In the case of the open boundary, the bulk-boundary correspondence states that if edge-localized modes exist, their energy spectrum is given by (\ref{eq:bbc1}).
Note that the edge modes are not guaranteed to appear within the open boundary condition.

We first consider the kagome lattice model.
We note that boundary modes exist for the ribbon geometry of this system illustrated in Fig.~\ref{fig:bbc}(a), which respects the translational symmetry along $(1/2,\sqrt{3}/2)$ while terminated along the $x$-axis.
The width $W$ of the kagome ribbon is defined as the number of the unit cells along the $x$-axis.
For example, the width of the kagome ribbon shown in Fig.~\ref{fig:bbc}(a) is 4.
We plot the band dispersions of the kagome ribbons with $W=20$ for $d_\mathrm{max}=1$ and $d_\mathrm{max}=0.8$ in Fig.~\ref{fig:bbc}(b) and (c), respectively.
The red and blue lines represent the boundary modes stemming from the band-crossing point at $k_y=0$.
While the band dispersions of the left- and right-localized modes are precisely the same for the $d_\mathrm{max}=1$ case, it is not for $0<d_\mathrm{max}<1$ case due to the broken time-reversal symmetry.
For this reason, we distinguish the left- and right-localized modes by the red and blue colors in Fig.~\ref{fig:bbc}(c).
We check that the blue and red curves, although they look asymmetric with respect to $k_y=0$, they follow the same parabolic equation (\ref{eq:bbc1}) in the vicinity of the touching-point at $k_y=0$.
We numerically calculate the effective mass of the boundary modes from the kagome lattice model and compare it with the analytic result of the effective mass $m^*=m_b/d^2_\mathrm{max}$ in (\ref{eq:bbc1}).
As plotted in Fig.~\ref{fig:bbc}(d), the formula (\ref{eq:bbc1}) describes the numerical results perfectly for any values of $d_\mathrm{max}$.
Second, we also investigate the edge state of the square lattice bilayer ribbon shown in Fig.~\ref{fig:bbc}(e).
As in the kagome model, the width $W$ of this system is defined as the number of unit cells along the $x$-axis.
In Fig.~\ref{fig:bbc}(f) and (g), we plot the band structures of the square lattice bilayer ribbon with $W=20$.
The red curves, which are doubly degenerate, correspond to the boundary modes.
We confirm that the effective mass of the boundary modes obeys the continuum formula (\ref{eq:bbc1}) well as plotted in Fig.~\ref{fig:bbc}(h).

\section{Conclusions}\label{sec:conclusions}
In summary, we propose a construction scheme for tight-binding Hamiltonians hosting a flat band whose band-touching point is characterized by $d_\mathrm{max}$, the maximum value of the quantum distance between Bloch eigenstates around the touching point.
Based on the scheme, we built several flat band tight-binding models with simple hopping structures in the kagome lattice and the square lattice bilayer, where one can control $d_\mathrm{max}$.
We note that complex and long-range (at least the next nearest ones) hopping amplitudes are necessary to change $d_\mathrm{max}$ between 0 and 1.
This implies that the candidate materials hosting a SFB with $0<d_\mathrm{max}<1$ could be found among the materials with strong spin-orbit coupling.
We believe that our construction scheme could inspire the material search for the geometrically nontrivial flat band systems.
If we extend the category of the materials to the artificial systems, our lattice models with the fine-tuned complex hopping parameters are expected to be realized in the synthetic dimensions~\cite{celi2014synthetic,ozawa2019topological,yuan2018synthetic,dutt2019experimental,balvcytis2022synthetic,ozawa2016synthetic,ozawa2017synthetic} and circuit lattices~\cite{albert2015topological,kim2023realization}.
Then, we propose a bulk-boundary correspondence between the bulk number $d_\mathrm{max}$ and the shape of the low-energy dispersion of the boundary modes within the open boundary condition.
The information of $d_\mathrm{max}$ is embedded in the effective mass of the band dispersion of the edge states.
This correspondence provides us with a tool to detect $d_\mathrm{max}$ from the spectroscopy of the finite SFB systems.
Notably, the bulk-boundary correspondence is obtained from the continuum Hamiltonian around the band-crossing point.
This implies that even if the flat band obtained from our construction scheme is slightly deformed in real systems, one can investigate the geometric properties of the singular flat band.
\\

\appendix
\section{Derivation of $\mathbf{d_{\rm max}}$ formula}\label{app: dmax}
Let us consider a linearized quantum state of the form
\ba
\alpha_\mathbf{k}\mathbf{v}_\mathbf{k} \approx q_1\mathbf{u}_1 + q_2\mathbf{u}_2,\label{eq:lin_apprx_cls}
\ea
where $\mathbf{u}_\mu $ is represented by a column vector of size equal to the number orbitals in a unit cell and $\alpha_\mathbf{k}$ is a factor introduced in Sec.~\ref{sec: Scheme}.
Here, $\mathbf{u}_1$ and $\mathbf{u}_2$ can take complex numbers as their components and do not have to be orthogonal with each other.
Without loss of generality, one can express $q_\mu$ as $q_1=q_x=q\cos\theta$ and $q_2 = q_x\sin\alpha + q_y\cos\alpha = q\sin(\theta + \alpha)$.
After normalization, we have
\ba
\mathbf{v}_\mathbf{k} \approx c(\alpha,\theta) \left( \cos\theta \mathbf{u}_1 + \sin(\theta+\alpha) \mathbf{u}_2 \right),
\ea
where
\ba
c(\alpha,\theta)=& \big[\mathbf{u}_1^*\cdot\mathbf{u}_1\cos^2(\theta+\alpha) + \mathbf{u}_1^*\cdot\mathbf{u}_2\cos(\theta+\alpha)\sin\theta  \nonumber\\
&+ \mathbf{u}_2^*\cdot\mathbf{u}_1\cos(\theta+\alpha)\sin\theta  + \mathbf{u}_2^*\cdot\mathbf{u}_2\sin^2\theta \big]^{-\frac{1}{2}}.
\ea
Then, the quantum distance between two states at $\theta_1$ and $\theta_2$ is given by
\begin{widetext}
	\ba
	d^2_\alpha(\theta_1,\theta_2)
	=1-\left| \frac{ \cos\theta_1 \left[\mathbf{u}_1^*\cdot\mathbf{u}_1\cos\theta_2 +  \mathbf{u}_1^*\cdot\mathbf{u}_2\sin(\theta_2+\alpha) \right] +\sin(\theta_1+\alpha) \left[ \mathbf{u}_2^*\cdot\mathbf{u}_2\cos\theta_2 +\mathbf{u}_2^*\cdot\mathbf{u}_2\sin(\theta_2+\alpha) \right]  }{c(\alpha,\theta_1) c(\alpha,\theta_2)} \right|^2.
	\ea
\end{widetext}
One can show that the maximum value of $d^2_\alpha(\theta_1,\theta_2)$ is independent of $\alpha$ and $\theta_1$.
Therefore, we assume that $\alpha=\theta_1=0$, which leads to
\ba
d^2_0(0,\theta)
=\frac{\left( ||\mathbf{u}_1||^2 ||\mathbf{u}_2||^2- |\mathbf{u}_1^*\cdot\mathbf{u}_2|^2 \right)\sin^2\theta }{ ||\mathbf{u}_1||^2 || \mathbf{u}_1\cos\theta + \mathbf{u}_2\sin\theta ||^2},
\ea
where $||\mathbf{v}||^2 = \mathbf{v}^*\cdot\mathbf{v}$.
From $dd^2_0(0,\theta)/d\theta = 0$, we obtain
\ba
\tan\theta_c = -2\frac{||\mathbf{u}_1||^2}{\mathbf{u}_1^*\cdot\mathbf{u}_2 + \mathbf{u}_2^*\cdot\mathbf{u}_1},
\ea
at which the quantum distance shows an extremum.
Then, the maximum quantum distance is evaluated as
\ba
d^2_{\rm max} =& d^2_0(0,\theta_c),\\
=& \frac{||\mathbf{u}_1||^2 ||\mathbf{u}_2||^2 -|\mathbf{u}_1^*\cdot\mathbf{u}_2|^2 }{||\mathbf{u}_1||^2 ||\mathbf{u}_2||^2 -(\mathrm{Re} \mathbf{u}_1^*\cdot\mathbf{u}_2)^2}, \\
=& \frac{1 -  |\mathbf{u}_1^*\cdot\mathbf{u}_2|^2}{ 1 - (\mathrm{Re} \mathbf{u}_1^*\cdot\mathbf{u}_2)^2},
\ea
where $\mathbf{u}_1$ and $\mathbf{u}_1$ are assumed to be normalized.

\section{A condition for the CLS to have a noninteger $d_\mathrm{max}$}\label{app:dmax_cond}

In this section, we show that at least one component of $\alpha_\mathbf{k}\mathbf{v}_\mathbf{k}$, the Fourier transform of the CLS, should contain more than two different exponential factors $e^{-i(mq_1+nq_2)}$.
Here, $q_i$ is the momentum with respect to the band-crossing point, and $m$ and $n$ are integer numbers.
To this end, we verify that if all the components of $\alpha_\mathbf{k}\mathbf{v}_\mathbf{k}$ have two or less than two exponential factors, $d_\mathrm{max}$ of the corresponding flat band is one or zero.
The $q$-th component of such an eigenvector can be written as
\ba
\alpha_\mathbf{k} \mathbf{v}_\mathbf{k}|_q = A_{m_1,n_1}e^{-i(m_1q_1+n_1q_2)} + A_{m_2,n_2}e^{-i(m_2q_1+n_2q_2)}.
\ea
%
Since we assume that the flat band is singular at the band-touching point, the coefficients satisfy
\ba
A_{m_1,n_1}+A_{m_2,n_2}=0.
\ea
As a result, the linear expansion of $\alpha_\mathbf{k}$ becomes
\ba
\alpha_\mathbf{k} \mathbf{v}_\mathbf{k}|_q \approx -iA_{m_1,n_1}\left[(m_1-m_2)q_1 -(n_1-n_2)q_1 \right],
\ea
leading to $u_{1,q}^*u_{2,q} = |A_{m_1,n_1}|^2(m_1-m_2)(n_1-n_2)$, where $u_{i,q}$ is the $q$-th component of $\mathbf{u}_i$ defined in (\ref{eq:lin_apprx_cls}).
Therefore $\mathbf{u}_1^*\cdot \mathbf{u}_2 = \sum_q u_{1,q}^*u_{2,q}$ is a real number, which proves the statement at the beginning of this section.
Namely, we need at least three different exponential factors in at least one component of $\alpha_\mathbf{k}$.

%

{\small \subsection*{Acknowledgements}
This work was supported by the National Research
Foundation of Korea (NRF) Grant funded by the Korea government
(MSIT) (Grant No. 2021R1A2C1010572 and 2021R1A5A1032996 and 2022M3H3A106307411).

\end{document}